\documentclass{article}

 \usepackage{hyperref}
\usepackage{bm}
\usepackage{amssymb}	
\usepackage{authblk}
\usepackage{graphicx}
\usepackage{tabularx, booktabs, makecell, caption}
\usepackage{siunitx}
\usepackage{float}\usepackage[
singlelinecheck=false 
]{caption}

\begin{document}
\title{\bf Thin accretion disk images of rotating hairy Horndeski black holes}
\author{{Mohaddese Heydari-Fard$^{1}$%
\thanks{Electronic address: \href{mailto:m\_heydarifard@sbu.ac.ir}{m\_heydarifard@sbu.ac.ir}} ,  Malihe Heydari-Fard$^{2}$\thanks{Electronic address: \href{mailto:heydarifard@qom.ac.ir}{heydarifard@qom.ac.ir}} and Nematollah Riazi$^{1}$\thanks{Electronic address: \href{mailto:n\_riazi@sbu.ac.ir}{n\_riazi@sbu.ac.ir}}}\\ {\small \emph{$^{1}$ Department of Physics, Shahid Beheshti University, Evin 19839, Tehran, Iran}}
\\{\small \emph{$^{2}$ Department of Physics, The University of Qom, 3716146611, Qom, Iran}}}

\maketitle

\begin{abstract}
By considering the steady-state Novikov-Thorne model, we study thin accretion disk processes for rotating hairy black holes in the framework of the Horndeski gravity. We obtain the electromagnetic properties of accretion disk around such black holes and investigate the effects of the hair parameter $h$ on them. We find that by decreasing the hair parameter from the Kerr limit, $h\rightarrow0$, the radius of the innermost stable circular orbit decreases which makes thin accretion disks around rotating hairy black holes in Horndeski gravity more efficient than that for the Kerr black hole in general relativity. Furthermore, using the numerical ray-tracing method, we plot thin accretion disk images around these black holes and investigate the effects of hair parameter on the central shadow area of accretion disk.
\vspace{5mm}
\vspace{1mm}
\end{abstract}

\section{Introduction}
Black holes (BHs) are solutions to Einstein field equations of general relativity (GR). Despite the historical debates on the BH’s existence, there have been several observational evidences in recent decades which support the existence of BHs. For instance, the discovery of gravitational waves from binary BH mergers \cite{GW}, the detection of stochastic gravitational wave background by the Pulsar Timing Array \cite{new1}--\cite{new2}and images of the shadow of BHs in the center of M87 galaxy and Milky Way \cite{sh1}--\cite{sh2}, are some of them. Moreover, the observed electromagnetic spectra of accretion disks around compact objects can also confirm the existence of BHs \cite{Yuan}--\cite{Bambi-1}.

It is well known that the astrophysical BHs gain their mass through the accretion process. In this process, the accreting matter accumulates into a BH by its gravitational attraction and releases gravitational energy in the form of radiation. The emission spectra of such radiation is associated to nearly geodesic motion of test particles and the structure of central BH. Therefore, the electromagnetic spectrum of accretion disk around a BH can be used to explore additional imprints of modified theory of gravity that we are interested in testing. Such studies including the investigation of accretion disks in $f(R)$ gravity, scalar-vector-tensor-gravity, brane-world scenario, Einstein-Maxwell-dilaton and Einstein-Gauss-Bonnet gravity have been done in \cite{FR1}--\cite{EGB4}. The effects of dark matter, non-linear electrodynamics and naked singularities on the disk properties have also been studied in \cite{d1}--\cite{d8}, respectively. Moreover, exploring the image of the accretion disk around a BH is another interesting topic that has been the focus of attention in recent years \cite{im-1}--\cite{im-7}.

Although the theory of GR has been well tested in weak and strong field regimes, it needs to be further modified to address its shortcomings, such as an inflationary era in the early universe, the nature of the dark matter and dark energy, the cosmological constant and the Hubble tension.

Among modifications of GR are the scalar-tensor theories in which one or more scalar fields couple to curvature and interfere with the gravitational interaction \cite{scalar}. The most general scalar-tensor theory in four-dimensions is the Horndeski gravity that leads to second-order field equations and it is free of Ostrogradski instabilities \cite{Horndeski}. Besides the cosmological studies in this theory \cite{c1}--\cite{de2}, the BH solutions of Horndeski gravity have been extensively studied \cite{b1}--\cite{b5}. Also, by considering the quartic scalar field model, the hairy BH solution of Horndeski gravity has been obtained in \cite{Bergliaffa}. The extension of this solution to the case of rotating hairy Horndeski BHs is done in \cite{Walia}. The authors have also investigated the gravitational deflection of light in such a space-time and by modelling of Sgr A* BH with hairy Horndeski BH showed that the correction in deflection angle can be measured with the current observational facilities. The strong lensing
observables of static hairy Horndeski BHs, including deflection angle, separation and magnification have been discussed in \cite{h1}. Testing Horndeski gravity from the observational data of the BH shadow has been considered in \cite{h2}--\cite{h3} and some constraints on the hair parameter have been obtained. The motion of test particles and their harmonic oscillations in the spacetime of static hairy BHs in Hordeski gravity, as well as the astrophysical applications of quasiperiodic oscillations has been studied in \cite{h4}. Furthermore, superradiant energy extraction \cite{h5}, perturbations of massless external fields \cite{h6}, and the effects of tilted thin accretion disks on the optical appearance of hairy Horndeki BHs \cite{h7} were studied.

The image of static hairy Horndeski BHs illuminated by thin disk and thin spherical accretion flows, has been investigated in \cite{h8}. Since the astrophysical BHs are expected to be rotating due to the accretion effects, in the present work we consider the rotating hairy Horndeski BHs and study thin accretion disk luminosity and its image around these BHs. We employ the steady-state Novikov-Thorne model \cite{Novikov} to study thin accretion disk processes of rotating BHs in Horndeski gravity, and investigate the effects of hair parameter on both the disk properties and its image.

The structure of the present paper is as follows. In section~\ref{2-Thin-disk}, we present the test particles motion and thin accretion disk model in a general stationary axisymmetric space-time. A brief review of rotating hairy Horndeski BHs is given in section~\ref{3-BH}. Then, the disk properties around such BHs are obtained in section~\ref{4-disk-properties} and the effects of the hair parameter are discussed. In section~\ref{5-disk-image}, we plot the ray-traced redshifted image and intensity and polarization profile of an accretion disk around a rotating hairy Horndeski BH. The paper ends with drawing our conclusions.

\section{Thin accretion disk model}
\label{2-Thin-disk}
In order to study the electromagnetic properties of accretion disk, we need to obtain the equations of motion of test particles in the space-time of the central compact object. Therefore, we take into account the Lagrangian ${\cal L}$ of a test particle in the vicinity of the central BH as follows
\begin{equation}
{\cal L}= \frac{1}{2}g_{\mu\nu}\dot{x^{\mu}}\dot{x^{\nu}},
\label{a1}
\end{equation}
where a dot denotes derivative with respect to the affine parameter and $g_{\mu\nu}$ is the line element of a generic stationary and axisymmetric
space-time
\begin{equation}
ds^2=g_{tt}dt^2+2g_{t\phi}dtd\phi+g_{rr}dr^2+g_{\theta\theta}d\theta^2+g_{\phi\phi}d\phi^2,
\label{a2}
\end{equation}
for which the metric components $g_{tt}, g_{rr}, g_{\theta\theta}$, $g_{\phi\phi}$ and $g_{t\phi}$ are assumed to be only functions of $r$ and $\theta$ coordinates. Thus, there are two constants of motion, the energy and the angular momentum per unit rest-mass, $\tilde{E}$ and $\tilde{L}$, which are given by
\begin{equation}
g_{tt}\dot{t}+g_{t\phi}\dot{\phi}=-\tilde{E},
\label{a3}
\end{equation}
\begin{equation}
g_{t\phi}\dot{t}+g_{\phi\phi}\dot{\phi}=\tilde{L}.
\label{a4}
\end{equation}
Using above equations and the normalization condition for massive particles, $g_{\mu\nu}\dot{x}^{\mu}\dot{x}^{\nu}=-1$, we obtain
\begin{equation}
g_{rr} \dot{r}^2+g_{\theta\theta}\dot{\theta}^2=V_{\rm eff}(r,\theta),
\label{a5}
\end{equation}
where the effective potential is
\begin{equation}
V_{\rm eff}(r,\theta)=-1+\frac{\tilde{E}^2g_{\phi\phi}+2\tilde{E}\tilde{L}g_{t\phi}+\tilde{L}^2g_{tt}}{g_{t\phi}^2-g_{tt}g_{\phi\phi}}.
\label{a6}
\end{equation}
Then, from the radial component of the geodesic equation with conditions $\dot{r}=\dot{\theta}=\ddot{r}=0$ for equatorial circular orbits, we find the following relation for the angular velocity, $\Omega=\dot{\phi}/\dot{t}$, \cite{book}
\begin{equation}
\Omega_{\pm}=\frac{-g_{t\phi,r}\pm\sqrt{(g_{t\phi,r})^2-g_{tt,r}g_{\phi\phi,r}}}{g_{\phi\phi,r}},
\label{a7}
\end{equation}
where the upper and lower signs refer to co-rotating and counter-rotating orbits, respectively. Now, using equations (\ref{a3})-(\ref{a4}) and from condition $g_{\mu\nu}\dot{x}^{\mu}\dot{x}^{\nu}=-1$, the specific energy ${\tilde{E}}$ and the specific angular momentum ${\tilde{L}}$  for a particle on a circular orbit can be expresed as
\begin{equation}
{\tilde{E}}=-\frac{g_{tt}+g_{t\phi}\Omega}{\sqrt{-g_{tt}-2g_{t\phi}\Omega-g_{\phi\phi}\Omega^2}},
\label{a8}
\end{equation}
\begin{equation}
{\tilde{L}}=\frac{g_{t\phi}+g_{\phi\phi}\Omega}{\sqrt{-g_{tt}-2g_{t\phi}\Omega-g_{\phi\phi}\Omega^2}}.
\label{a9}
\end{equation}
The innermost stable circular orbit, $r_{\rm isco}$, can be obtained from the condition $V_{\rm eff,rr}=0$, which leads to the following equation
\begin{equation}
V_{\rm eff,rr}\mid_{r=r_{\rm isco}}=\frac{1}{g_{t\phi}^2-g_{tt}g_{t\phi}}\left[{\tilde{E}^2g_{\phi\phi,rr}}
+2\tilde{E}\tilde{L}g_{t\phi,rr}+{\tilde{L}^2g_{tt,rr}}-\left(g_{t\phi}^2-g_{tt}g_{\phi\phi}\right)_{,rr}\right]\mid_{r=r_{\rm isco}}=0.
\label{a10}
\end{equation}

The standard framework to describe thin accretion disk processes is the Novikov-Thorne model \cite{Novikov}, which is a generalization of the Newtonian approach of Shakura-Sunyaev \cite{Shakura}. The model is based on some typical assumptions such as the vertical size of disk, $H$, is negligible compared to its horizontal size, $H\ll r$. Also, it is assumed that the disk is in the equatorial plane of the central compact object and its inner edge is located at the ISCO radius. The disk is considered in a steady-state which means that the mass accretion rate, $\dot{M}$, is constant in time. In this steady-state model the accreting gas is assumed to be in thermodynamical equilibrium and thus the disk temperature is related to the energy flux via the equation
\begin{equation}
F(r)=\sigma_{\rm SB} T^4(r),
\label{a11}
\end{equation}
where $\sigma_{\rm SB}=5.67\times10^{-5}\rm erg$ $\rm s^{-1} cm^{-2} K^{-4}$ is the Stefan-Boltzmann constant and the energy flux emitted from the disk surface is given by \cite{Novikov,Page}
\begin{equation}
F(r)=-\frac{\dot{M}_{0}\Omega_{,r}}{4\pi\sqrt{-g}\left(\tilde{E}-\Omega \tilde{L}\right)^2}\int^r_{r_{\rm isco}}\left(\tilde{E}-\Omega \tilde{L}\right) \tilde{L}_{,r}dr.
\label{a12}
\end{equation}
Moreover, the luminosity of disk can be found as \cite{Torres}
\begin{equation}
L(\nu)=4\pi d^2 I(\nu)=\frac{16\pi^2\hbar\cos\gamma}{c^2}\int_{r_{\rm in}}^{r_{\rm out}}\int_0^{2\pi}\frac{ \nu_{\rm e}^3 r dr d\phi}{\exp{[\frac{h\nu_{\rm e}}{k_{\rm B} T}]}-1
},\label{a13}
\end{equation}
where $d$ is the distance from the disk center, $\gamma$ is the disk inclination angle, and $r_{\rm in}$ and $r_{\rm out}$ denote the inner and outer radii of the edge of disk, respectively. Also, $\hbar$ and $k_{\rm B}$ are respectively the reduced Planck constant and Boltzmann constant, while $\nu_{\rm e}=\nu(1+z)$ is the emitted frequency with the redshift factor $z$ given by
\begin{equation}
1+z=\frac{1+\Omega r\sin\phi\sin\gamma}{\sqrt{-g_{tt}-2g_{t\phi}\Omega-g_{\phi\phi}\Omega^2}}.
\label{a14}
\end{equation}
Finally, we define the accretion efficiency as
\begin{equation}
\epsilon=1-\tilde{E}_{\rm isco}.
\end{equation}
which denotes the capability of central massive object to convert rest mass into radiation.

\section{Rotating BHs in Horndeski gravity}
\label{3-BH}
The action of Horndeski gravity is described by
\begin{equation}
{\cal S}=\int d^4x \sqrt{-g}\left[Q_2(\chi)+Q_3(\chi)\Box\phi+Q_4(\chi)R+Q_{4,\chi}\left((\Box\phi)^2-(\nabla^{\mu}\nabla^{\nu}\phi)(\nabla_{\mu}\nabla_{\nu}\phi)\right)\right],
\label{h1}
\end{equation}
where $g$, $R$ and $\phi$ are the metric determinant, Ricci scalar and a scalar field, respectively. Also, $Q_{i=2,..4}(\chi)$ are arbitrary functions of the kinetic term $\chi=-1/2\partial^{\mu}\phi\partial_{\mu}\phi$. Note the above action is a quartic Horndeski scalar field model in which the terms containing $Q_5$ are absent \cite{Bergliaffa}. From the definition of the four-current
\begin{equation}
j^{\nu}=\frac{1}{\sqrt{-g}}\frac{\delta{\cal S}}{\delta(\phi_{,\nu})},
\label{h02}
\end{equation}
it follows that
\begin{eqnarray}
j^{\nu}&=& -Q_{2,\chi}\phi^{,\nu}-Q_{3,\chi}(\phi^{,\nu}\Box\phi+\chi^{,\nu})-Q_{4,\chi}(\phi^{,\nu}R-2R^{\nu\sigma}\phi_{,\sigma})\nonumber\\
&-&Q_{4,\chi,\chi}[\phi^{,\nu}((\Box\phi)^2-(\nabla_{\alpha}\nabla_{\beta}\phi)(\nabla^{\alpha}\nabla^{\beta}\phi))+2(\chi^{,\nu}\Box\phi-\chi_{,\mu}\nabla^{\mu}\nabla^{\nu}\phi)],
\label{h2}
\end{eqnarray}
where we have used the usual convention of the Riemann tensor
\begin{equation}
\nabla_{\rho}\nabla_{\beta}\nabla_{\alpha}\phi-\nabla{\beta}\nabla{\rho}\nabla_{\alpha}\phi=-R^{\sigma}_{\alpha\rho\beta}\nabla_{\sigma}\phi.
\label{h002}
\end{equation}
Varying action (\ref{h1}) with respect to metric $g^{\mu\nu}$ we find the field equations
\begin{equation}
Q_4G_{\mu\nu}=T_{\mu\nu},
\label{h3}
\end{equation}
where $T_{\mu\nu}$ is given by
\begin{eqnarray}
T_{\mu\nu}&=&\frac{1}{2}(Q_{2,\chi}\phi_{,\mu}\phi_{,\nu}+Q_2g_{\mu\nu})+\frac{1}{2}Q_{3,\chi}(\phi_{,\mu}\phi_{,\nu}\Box\phi-g_{\mu\nu}\chi_{,\alpha}\phi^{,\alpha}
+\chi_{,\mu}\phi_{\nu}+\chi_{,\nu}\phi_{\mu})\nonumber\\
&-&Q_{4,\chi}\left(\frac{1}{2}g_{\mu\nu}[(\Box\phi)^2-(\nabla_{\alpha}\nabla_{\beta}\phi)(\nabla^{\alpha}\nabla^{\beta}\phi)-2R_{\sigma\gamma}\phi^{,\sigma}\phi^{,\gamma}]
-\nabla_{\mu}\nabla_{\nu}\phi\Box\phi\right.\nonumber\\
&+&\left.\nabla_{\gamma}\nabla_{\mu}\phi\nabla^{\gamma}\nabla_{\nu}\phi-\frac{1}{2}\phi_{,\mu}\phi_{,\nu}R+R_{\sigma\mu}\phi^{,\sigma}\phi_{,\nu}
+R_{\sigma\nu}\phi^{,\sigma}\phi_{,\mu}+R_{\sigma\nu\gamma\mu}\phi^{,\sigma}\phi^{,\gamma}\right)\nonumber\\
&-&Q_{4,\chi,\chi}\left(g_{\mu\nu}(\chi_{\alpha}\phi^{,\alpha}\Box\phi+\chi_{\alpha}\chi^{,\alpha})+\frac{1}{2}\phi_{,\mu}\phi_{,\nu}(
\nabla_{\alpha}\nabla_{\beta}\phi\nabla^{\alpha}\nabla^{\beta}\phi-(\Box\phi)^2)\right.\nonumber\\
&-&\left.\chi_{,\mu}\chi_{,\nu}-\Box\phi(\chi_{,\mu}\phi_{,\nu}+\chi_{,\nu}\phi_{,\mu})-\chi_{,\gamma}[\phi^{,\gamma}\nabla_{\mu}\nabla_{\nu}\phi
-(\nabla^{\gamma}\nabla_{\mu}\phi)\phi_{,\nu}-(\nabla^{\gamma}\nabla_{\nu}\phi)\phi_{,\mu}]\right).
\label{h4}
\end{eqnarray}

\begin{figure}[H]
\centering
\includegraphics[width=3.0in]{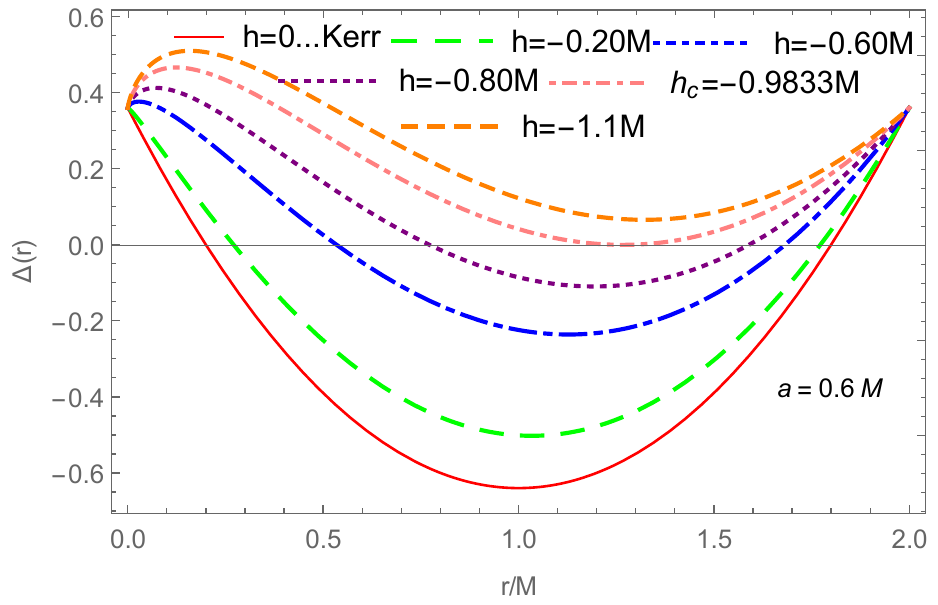}
\includegraphics[width=3.0in]{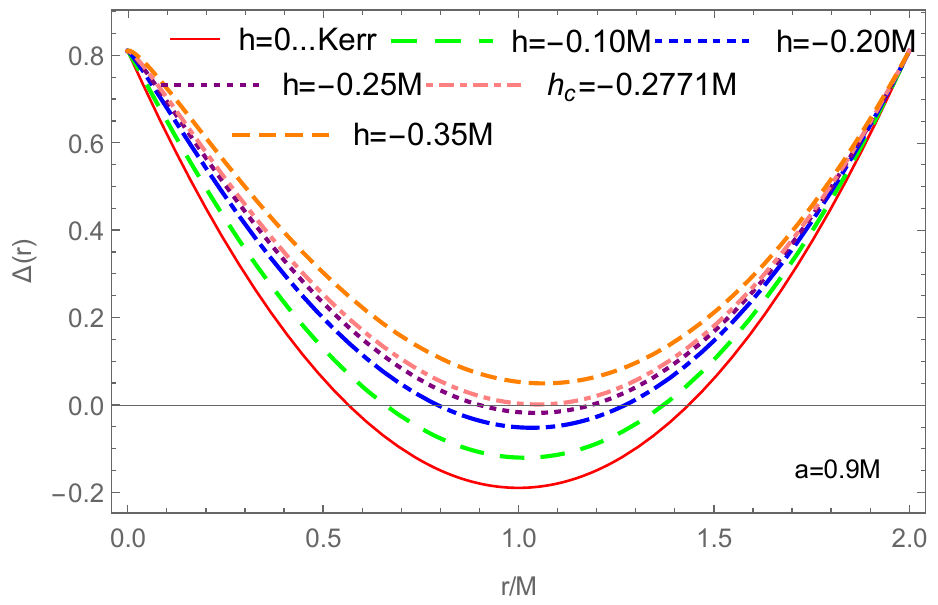}
\caption{\footnotesize The behavior of horizons as a function of the radial coordinate $r$ for different values of the hair parameter $h$. The rotation parameter is set to $a=0.6M$ (left panel) and $a=0.9M$ (right panel), respectively. In each panel the solid curve corresponds to the Kerr BH.}
\label{Delta}
\end{figure}

By considering the canonical action for the scalar field $\phi=\phi(r)$, and also imposing conditions of the finite energy of $\phi$, i.e., $E=\int_{V}\sqrt{-g}T^0_0d^3x$ and a vanishing radial four-current at infinity $j^r=0$, one can solve the field equations (\ref{h3}) and obtain the static and spherically symmetric solution as
\begin{equation}
ds^2=-f(r)dt^2+\frac{dr^2}{f(r)}+r^2(d\theta^2+\sin^2\theta d\varphi^2),
\label{h5}
\end{equation}
with
\begin{equation}
f(r) = 1-\frac{2M}{r}+\frac{h}{r}\ln\left(\frac{r}{2M}\right),
\label{h6}
\end{equation}
where $M$ is the BH mass and $h$ is referred to as the hair parameter with the dimension of length. When $h/M>0$ there is only one horizon at $r=2M$, while for the values of $h$ in the interval $-2<h/M<0$, there are two horizons; one is the outer event horizon at $r_+=2M$ and other is the inner Cauchy horizon, $r_{-}$. These two horizons coincide at $r=2M$ and we get an extremal BH \cite{Bergliaffa}. Also, when $h\rightarrow0$ the metric (\ref{h6}) reduces to the Schwarzschild solution, and it is asymptotically flat. Thus, here we restrict our analysis to the case of $-2<h/M<0$ as considered in \cite{Walia}--\cite{h8}. Using the revised Newman-Janis algorithm, Walia et al., obtained the rotating BH in Horndeski gravity which in the Boyer–Lindqist coordinates is given by \cite{Walia}
\begin{eqnarray}
ds^2&=&-\left(\frac{\Delta-a^2\sin^2\theta}{\Sigma}\right)dt^2+\frac{\Sigma}{\Delta}{dr^2}+\Sigma d\theta^2\nonumber\\
&+&\frac{2a\sin^2\theta}{\Sigma}\left(\Delta-(r^2+a^2)\right)dt d\varphi\nonumber\\
&+&\frac{\sin^2\theta}{\Sigma}\left[(r^2+a^2)^2-\Delta a^2\sin^2\theta\right]d\varphi^2,
\label{h7}
\end{eqnarray}
with
\begin{equation}
\Delta=r^2+a^2-2Mr+hr\ln\left(\frac{r}{2M}\right), \quad \Sigma=r^2+a^2\cos^2\theta.
\label{h8}
\end{equation}
In the absence of hair, $h\rightarrow0$, the above metric reduces to the Kerr metric and for $a=0$ it reduces to the line element of a static hairy Horndeski BH in equation (\ref{h5}). From the condition $g^{rr}=\Delta=0$ one can obtain the location of horizons. We have plotted $\Delta(r)$ as a function of radial coordinate for different values of $h$, and for $a=0.6M$ and $a=0.9M$ in Fig.~\ref{Delta}. As is clear, in each panel for some values of the hair parameter $h$, there are two horizons, namely the inner Cauchy horizon, $r_-$, and the outer event horizon, $r_+$, so that $r_-<r_+$. However for a critical value of $h=h_{\rm c}$, the two horizons coincide, $r_-=r_+$, corresponding to an extremal BH. Also, for $h<h_{\rm c}$ there are no BH space-time \cite{h2}. Also, we see that the radius of the event horizon decrease with increasing $|h|$.

\section{Thin accretion disk properties of rotating hairy Horndeski BHs}
\label{4-disk-properties}
\begin{figure}[H]
\centering
\includegraphics[width=3.0in]{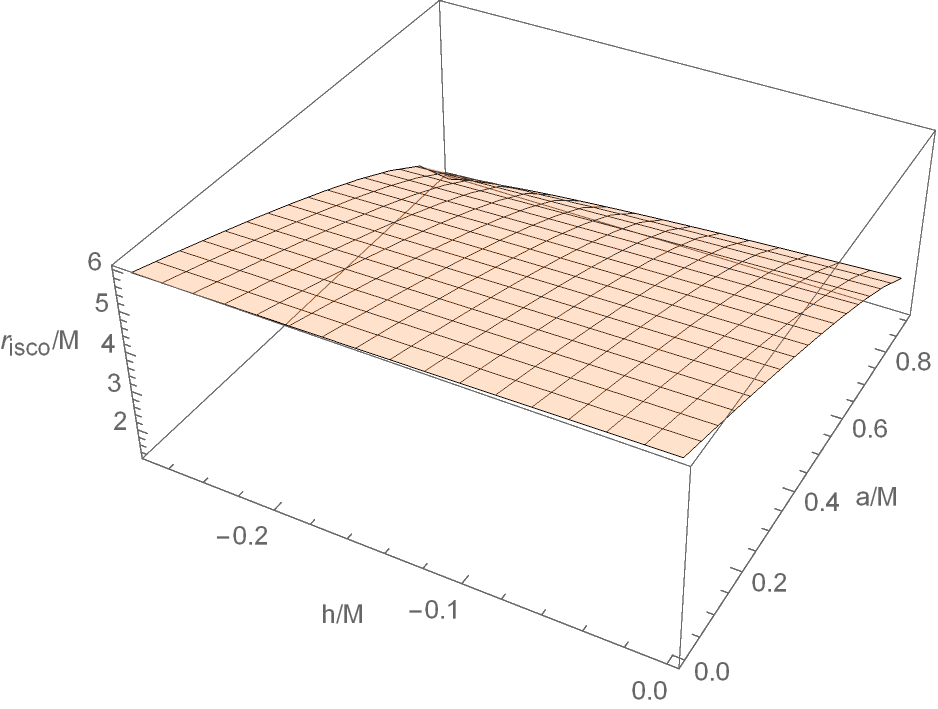}
\caption{\footnotesize The behaviour of the ISCO radius as a function of Horndeski parameter $h$ and the spin parameter $a$.}
\label{isco}
\end{figure}

Let us now study thin accretion disk properties around rotating hairy BHs in Horndeski gravity and compare the results with that of the Kerr BH in GR. First, using equations
(\ref{a7})-(\ref{a9}) we obtain the angular velocity, specific energy and specific angular momentum of test particles moving in circular orbits around a rotating hairy Horndeski BH as follows
\begin{equation}
\Omega=\frac{1}{a+\sqrt{2}r/u(r)},
\label{d1}
\end{equation}
\begin{equation}
{\tilde{E}}=\frac{\sqrt{2}(r-2M+h\ln(\frac{r}{2M}))+au(r)}{\sqrt{r[-6M+2r-h(1-3\ln(\frac{r}{2M}))+2\sqrt{2}au(r)]}},
\label{d2}
\end{equation}
\begin{equation}
{\tilde{L}}=\frac{-\sqrt{2}a(2M-h\ln(\frac{r}{2M}))+(r^2+a^2)u(r)}{\sqrt{r[-6M+2r-h(1-3\ln(\frac{r}{2M}))+2\sqrt{2}au(r)]}},
\label{d3}
\end{equation}
with
\begin{equation}
u(r)=\sqrt{\frac{2M+h(1-\ln(\frac{r}{2M}))}{r}}.
\label{d4}
\end{equation}
The innermost stable circular orbit (ISCO) equation (\ref{a10}) also reads
\begin{eqnarray}
&&2r(6M^2-Mr+4Mh+h^2)-4\sqrt{2}ar(2M+h)u(r)\nonumber\\
&+&3rh^2\ln(\frac{r}{2M})^2+a^2(6M+4h)+h\ln(\frac{r}{2M})\nonumber\\
&\times&\left(-3a^2+r(r-12M-4h)+4\sqrt{2}aru(r)\right)=0,
\label{d5}
\end{eqnarray}
which can not be solved analytically. Therefore, we have numerically obtained the ISCO radius for different values of $a$ and $h$ parameters. Also, in the absence of hair, $h=0$, with $a=0$ the above equation reduces to $r_{\rm isco}=6M$, which is the ISCO radius of the Schwarzschild BH. In Fig.~\ref{isco}, we have shown the effects of $a$ and $h$ parameters on the ISCO radius, showing that with increasing both the spin parameter and the absolute value of the hair parameter the ISCO radius decreases.

\begin{figure}[H]
\centering
\includegraphics[width=3.0in]{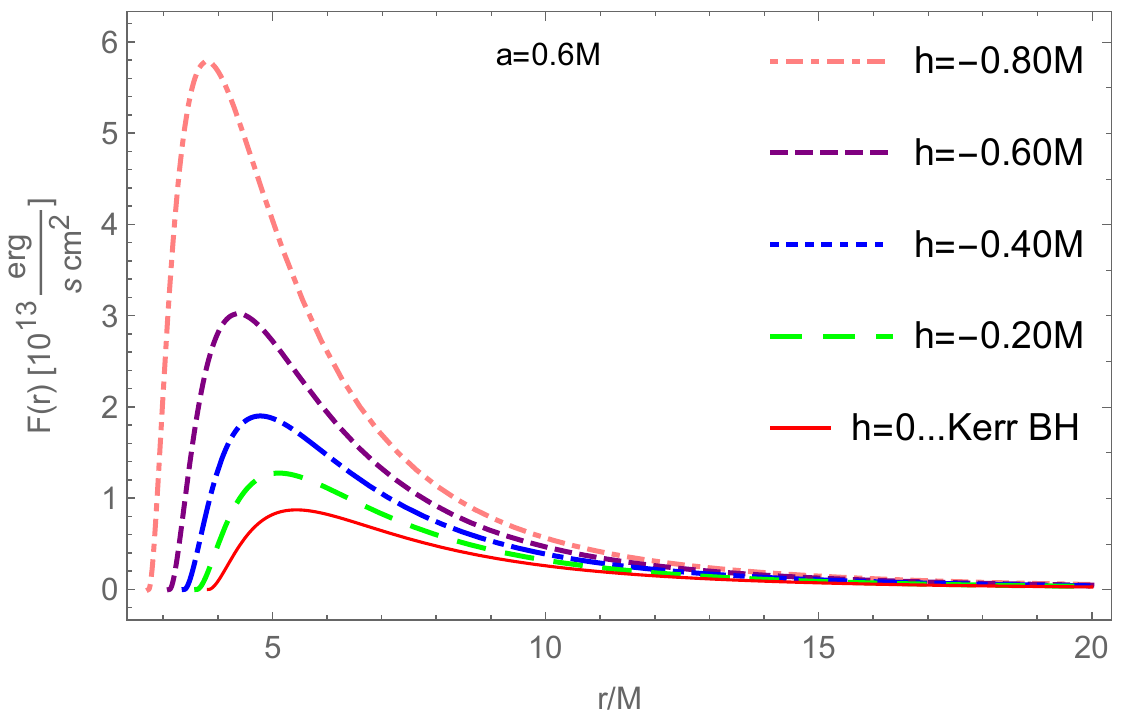}
\includegraphics[width=3.0in]{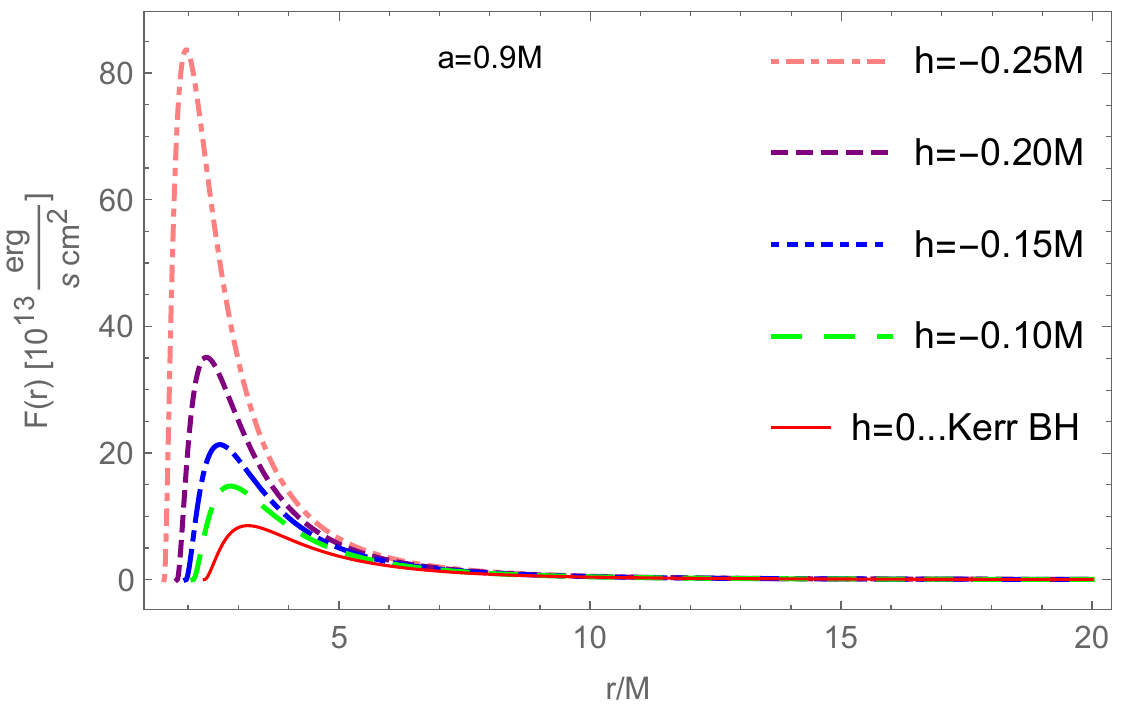}
\caption{\footnotesize The energy flux $F(r)$ from a disk around a rotating hairy Horndeski BH with the mass accretion rate $\dot{M}=2\times10^{-6}M_{\odot}\rm yr^{-1}$, for different values of the hair parameter $h$. The rotation parameter is set to $a=0.6M$ (left panel) and $a=0.9M$ (right panel), respectively. In each panel the solid curve corresponds to the Kerr BH.}
\label{flux}
\end{figure}

\begin{figure}[H]
\centering
\includegraphics[width=3.0in]{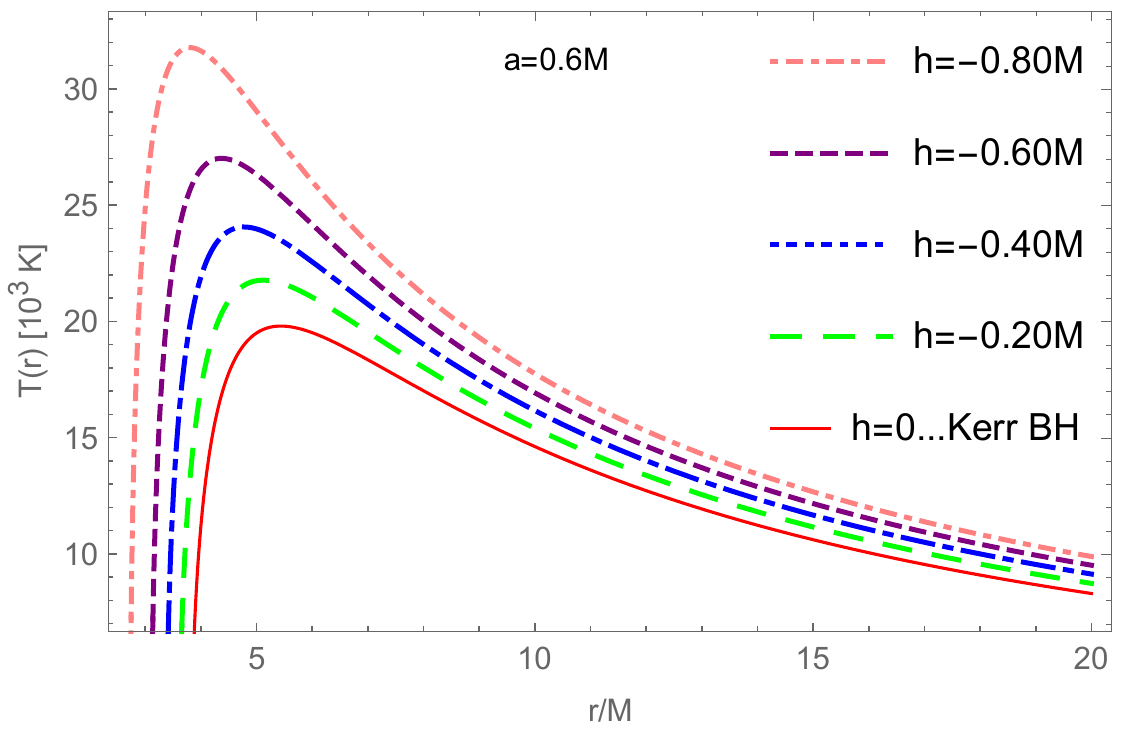}
\includegraphics[width=3.0in]{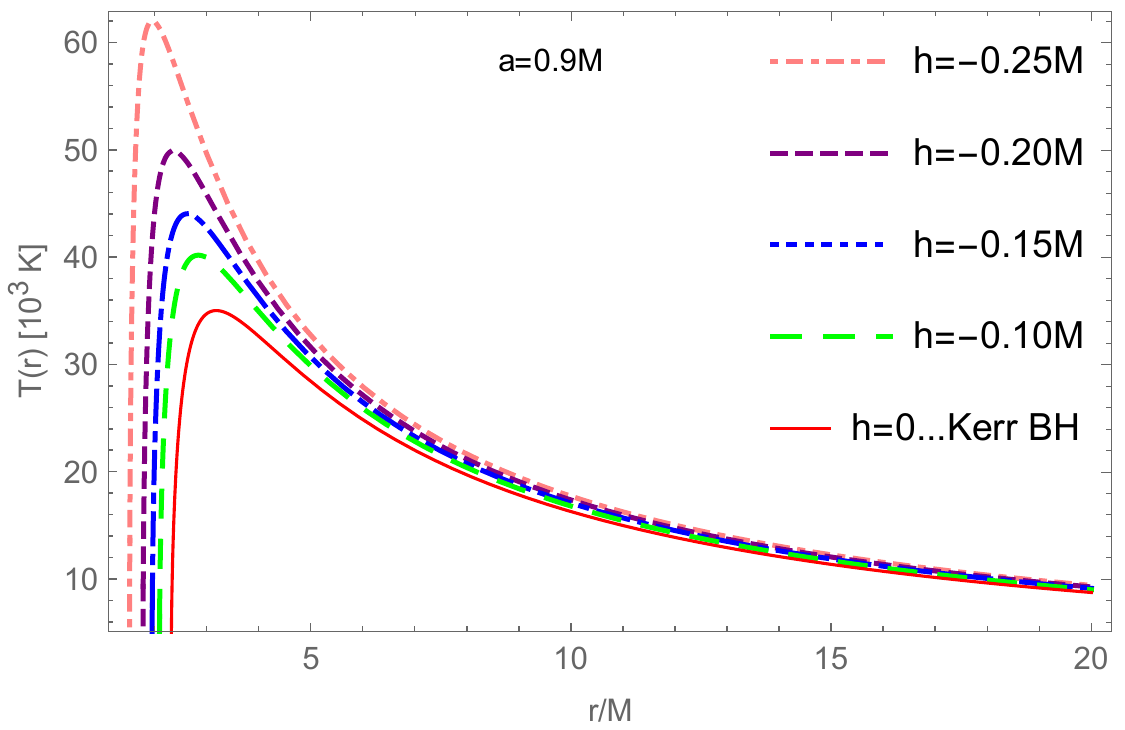}
\caption{\footnotesize The disk temperature $T(r)$ for a rotating hairy Horndeski BH with mass accretion rate $\dot{M}=2\times10^{-6}M_{\odot}\rm yr^{-1}$, for different values of the hair parameter $h$. The rotation parameter is set to $a=0.6M$ (left panel) and $a=0.9M$ (right panel), respectively. In each panel the solid curve corresponds to the Kerr BH.}
\label{temperature}
\end{figure}

Now, using equations (\ref{d1})-(\ref{d3}) for the space-time of rotating hairy Horndeski BHs, we are ready to investigate the effects of hair as well as the spin parameter on the electromagnetic properties of thin disk around these BHs. The behavior of the energy flux for $a=0.6M$ and $a=0.9M$ are shown in the left and right panels of Fig.~\ref{flux}, respectively. As we see from Fig.~\ref{Delta}, the critical value of hair parameter for $a=0.6M$ is $h_{\rm c}=-0.9833M$ and in the case of $a=0.9M$, is $h_{\rm c}=-0.2771M$. So, in each panel, the $h$ values were chosen in such a way that are larger than the critical value $h_{\rm c}$ and thus we have a BH solution. We see that for a fixed $a$, with increasing $|h|$, the radiant energy flux increases too, so that the Kerr BH with $h=0$ has the smallest energy flux. Also, an enhancement in the rotation parameter increases the energy flux. The disk temperature is plotted in Fig.~\ref{temperature}, where a similar behavior can be seen.

\begin{figure}[H]
\centering
\includegraphics[width=3.0in]{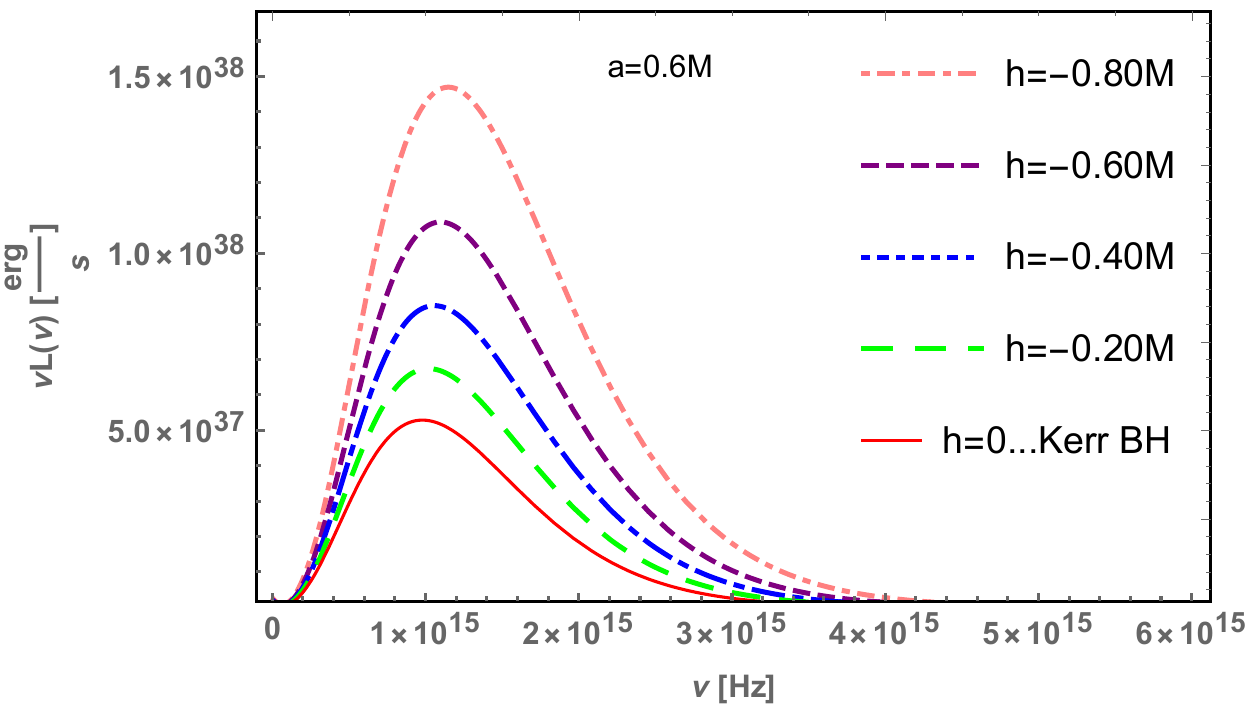}
\includegraphics[width=3.0in]{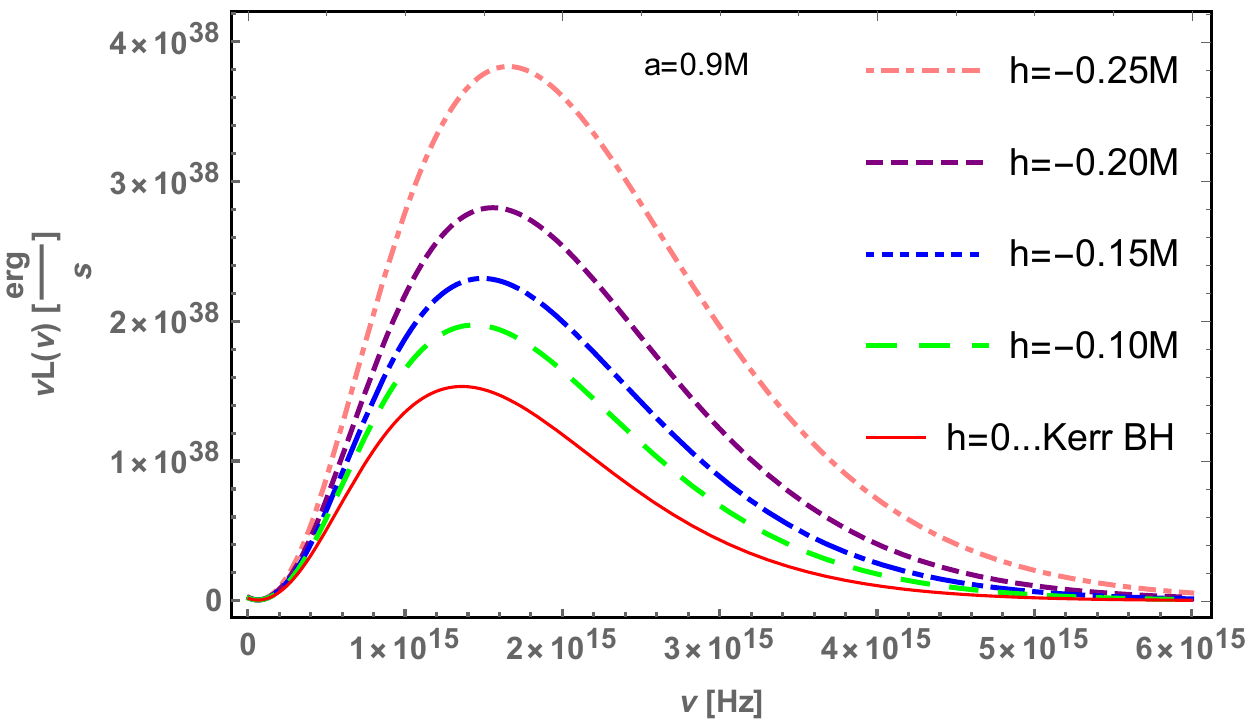}
\caption{\footnotesize The emission spectrum $\nu L(\nu)$ of the accretion disk around a rotating hairy Horndeski BH with mass accretion rate $\dot{M}=2\times10^{-6}M_{\odot}yr^{-1}$ and inclination $\gamma=0^{\circ}$ for different values of the hair parameter $h$, as a function of frequency $\nu$. The rotation parameter is set to $a=0.6M$ (left panel) and $a=0.9M$ (right panel), respectively. In each panel the solid curve corresponds to the Kerr BH.}
\label{luminosity}
\end{figure}

In Fig.~\ref{luminosity},we have displayed the effect of the hair parameter on the disk spectra for rotating Horndeski BHs. Similar to the energy flux and the disk temperature, deviation of $L(\nu)$ for Horndeski BHs from that of the Kerr BH becomes more pronounced with increasing $|h|$. Moreover, the cut-off frequencies, in which the maximum of luminosity occur, shift to higher frequencies with increase of $|h|$.

\begin{figure}[H]
\centering
\includegraphics[width=3.0in]{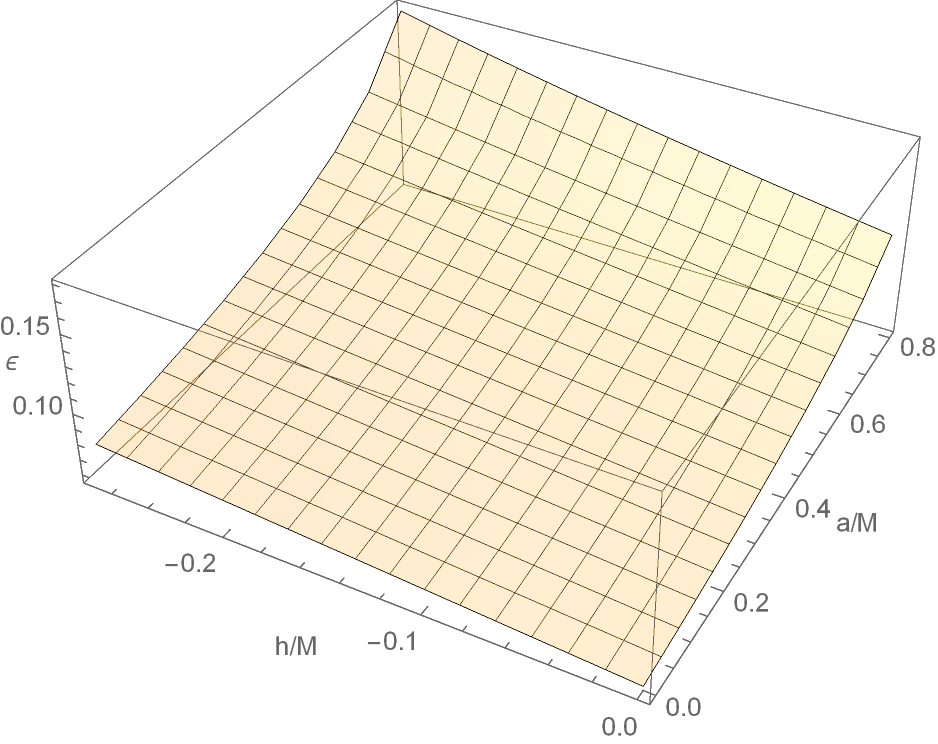}
\caption{\footnotesize The accretion efficiency $\epsilon$ of the Horndeski BHs as a function of the hair parameter $h$ and the spin parameter $a$. }
\label{efficiency}
\end{figure}

The dependence of the accretion efficiency $\epsilon$ on both the spin and hair parameter is plotted in Fig.~\ref{efficiency}. It shows that the radiative efficiency of Horndeski BHs increases  with increasing of $a$ and $|h|$. It means that the accretion of matter on hairy Horndeski BHs is more efficient than that on the Kerr BH with $h=0$. Therefore, the rotating hairy Horndeski BHs can provide a more efficient processes for converting the energy of accreting matter into electromagnetic radiation than that the Kerr BH in the absence of hair.

\section{Image of disk around a rotating hairy Horndeski BH}
\label{5-disk-image}
\begin{figure}
\centering
\includegraphics[width=3.2in]{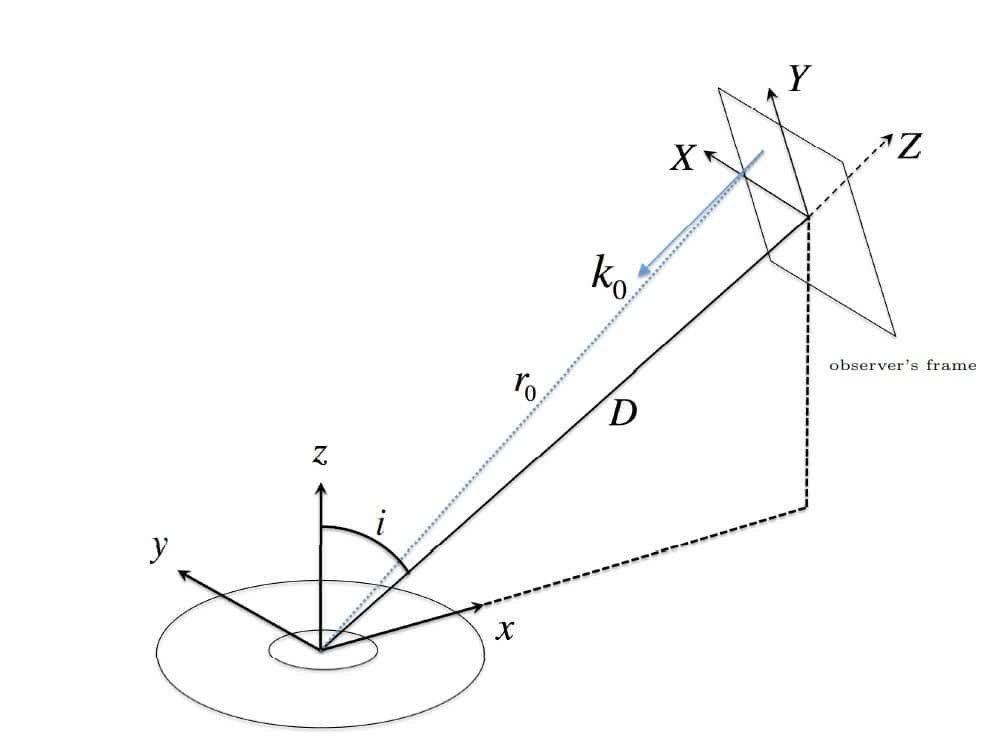}\par
\caption{\footnotesize Cartesian coordinates (x,y,z) are centred at the BH, and the Cartesian coordinates (X,Y,Z) are for the image plane of distant observer, located at the distant $D$ from the BH with inclination angle $i$. Figure from \cite{Bambi}.}
\label{coordinate}
\end{figure}

In this section, we use the numerical ray tracing method proposed by Chen et al \cite{Chen} to construct the image of a thin accretion disk around a rotating hairy Horndeski BH. The coordinate system of the BH and its surrounding disk and also the coordinate system of the distant observer is shown in Fig.~\ref{coordinate} and has been discussed in details in Ref. \cite{Bambi}.

The BH surrounded by an accretion disk is described with Cartesian coordinates system $(x,y,z)$, while the observer at a distant $D$ from the BH with inclination angle $i$ is described with Cartesian coordinates system $(X,Y,Z)$. Here, the image plane lies in the plane $Z=0$, meaning that the $Z$-axis is perpendicular to the image plane. The relation between two coordinates is given by
\begin{eqnarray}
&x = D \sin i-Y\cos i+Z \sin i,\nonumber\\
& y = X,\nonumber\\
&z = D\cos i+Y \sin i+Z \cos i.
\label{I1}
\end{eqnarray}

\subsection{Photon initial conditions}
For the first step we write the photon initial conditions in the image plane of the distant observer. For this purpose we consider the photon places at the image plane at the position $(X_0,Y_0,0)$ with 4-momentum ${k}^{\alpha}=(k_0,0,0,-k_0)$ so that the photon 3-momentum $\mathbf{k_0}=-k_0\hat{Z}$ is perpendicular to the image plane. The initial conditions for the photon position are given as
\begin{eqnarray}
&t_0 = 0,\nonumber\\
&r_0 = \sqrt{D^2+X_0^2+Y_0^2},\nonumber\\
& \theta_0 = \arccos \frac{D \cos i+ Y_0 \sin i}{r_0},\nonumber\\
&\phi_0 = \arctan\frac{X_0}{D\sin i-Y_0\cos i}.
\label{I2}
\end{eqnarray}

Using the relation $k^{\mu} = (\partial x^{\mu}/\partial \tilde{x}^{\alpha})\tilde{k}^{\alpha}$, where $\tilde{k}^{\mu}$ is the photon 4-momentum in Cartesian coordinates, we obtain the photon 4-momentum in the spherical coordinates system
\begin{eqnarray}
&k_0^r = -\frac{D}{r_0}|\mathbf{k_0}|,\nonumber\\
&k_0^\theta = \frac{\cos i-(Y_0 \sin i+D\cos i)\frac{D}{r_0^2}}{\sqrt{X_0^2+(D\sin i-Y_0\cos i)^2}} |\mathbf{k_0}|,\nonumber\\
&k_0^\phi = \frac{X_0 \sin i}{X_0^2+(D\sin i-Y_0\cos i)^2} |\mathbf{k_0}|,
\label{I3}
\end{eqnarray}
where $k_0^t$ cab be obtained from the condition $g_{\mu\nu} k^{\mu} k^{\nu} = 0$ as follows
\begin{eqnarray}
k_0^t = \sqrt{(k_0^r)^2+r_0^2(k_0^{\theta})^2+r_0^2\sin^2\theta_0(k_0^{\phi})^2}.
\label{I4}
\end{eqnarray}

\subsection{Photon trajectory}
Now by using the initial conditions (\ref{I2})-(\ref{I4}) we can integrate the following geodesic equation
\begin{equation}
\frac{d^2x^{\mu}}{d\lambda^2}+\Gamma^{\mu}_{\nu\rho}\frac{dx^{\nu}}{d\lambda}\frac{dx^{\rho}}{d\lambda}=0,
\label{I5}
\end{equation}
backward in time from detection point $(X_0,Y_0,0)$ at the image plane of the distant observer to the emission point. In equation (\ref{I5}), $\lambda$ is the affine parameter. In a similar way with Ref. \cite{Chen} we first solve the geodesic equations using the Runge-Kutta method in MATLAB software.

\subsection{Strong lensing by hairy Horndeski BH}
It is believed that the large luminosity of active galactic nuclei (AGN) and quasars is the result of gas accreted by supersessive BH. The quasar accretion disk emits X-rays. This emission is lensed by the central BH before it arrives at distant observer. The radiation intensity profile emitted from the disk surface has been observed to follow a power law. In this work, following \cite{Chen}, we consider the radiation intensity profile emitted from the disk surface as
\begin{eqnarray}
I(\nu,\mu,r)\propto\frac{1}{r^n}\frac{\omega(\mu)}{\nu^{\Gamma-1}},
\label{I6}
\end{eqnarray}
where $\nu$ is the frequency of photon in the rest frame and $\mu$ is the cosine between photon 4-momentum and the normal vector of disk measured by the comoving observer. $\omega(\mu)$ denotes the angular-dependence of intensity profile and we take it from Chandrasekhar's book \cite{Cbook}. Also, $\Gamma$ and $n$ are the photon index and the radial steepness of the profile, respectively. For the source profile (\ref{I6}) the following flux integral
\begin{eqnarray}
F_{\nu_{\rm o}} \equiv \int g^3 I^{\rm source}_{\nu_{\rm e}}(e^a,p_b) d\Omega_{0},
\label{I7}
\end{eqnarray}
can be easily computed analytically or numerically through backward ray-tracing. Here ${\nu_{\rm e}}$ is the source frequency and $g\equiv \frac{{\nu_{\rm o}}}{{\nu_{\rm e}}}$ is the redshift factor.

In Ref. \cite{Chen} authors investigated the polarization and gravitational Faraday rotation for rotating Kerr BH in details. For further study, the interested reader is referred  to \cite{B}--\cite{S}. Note that for a rotating BH, the photon polarization plane rotates due to gravitational Faraday rotation \cite{Connors}, which makes the disk radiation being partially polarized due to the magnetic field of the disk atmosphere. Also, due to the Thomson scattering of photons by free electrons in the dense atmosphere of accretion disk, the degree of polarization varies depending on the change in the disk’s inclination angle, so that as it increases the degree of polarization increases too.

\begin{figure}[H]
\centering
\includegraphics[width=3.0in]{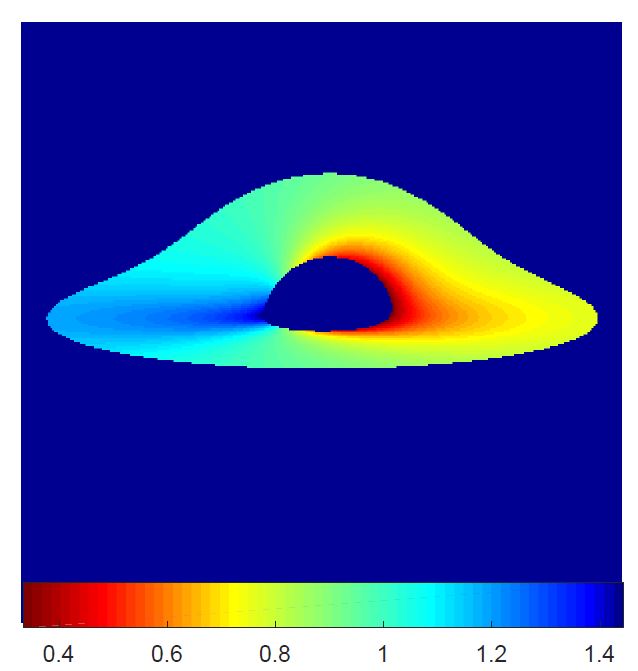}
\includegraphics[width=3.0in]{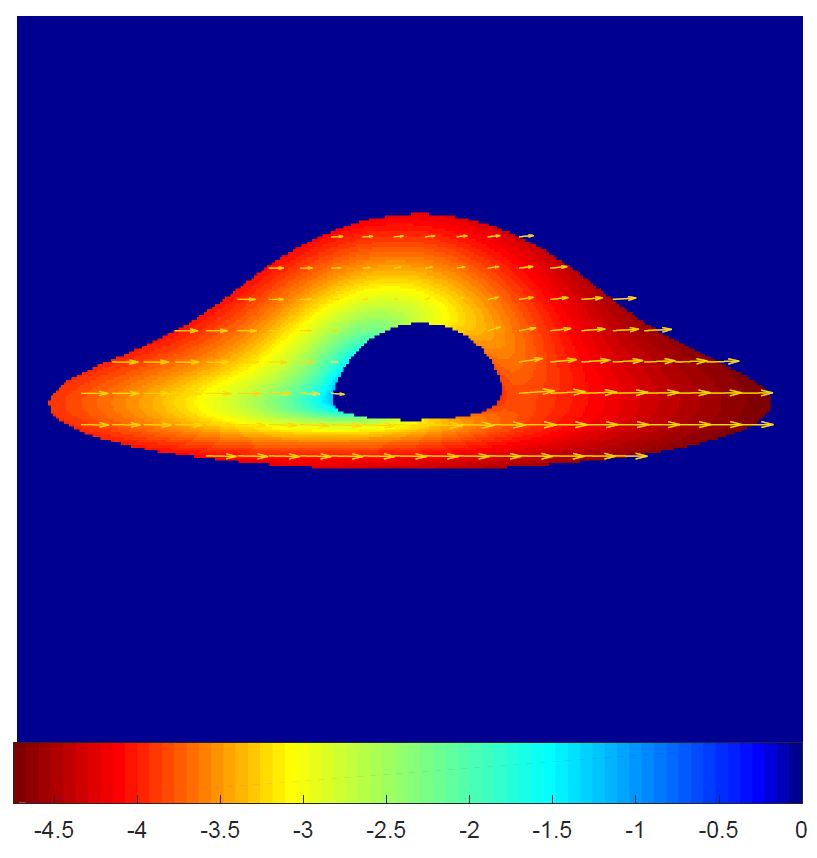}
\caption{\footnotesize Ray-traced redshifted image (left panel) and intensity and polarization profile (right panel) of a lensed accretion disk around Kerr BH. The inclination angle is set to $i=80^{\circ}$ and spin parameter is $a=0.6M$ with $M=1$. The inner edge of disk is at $r_{\rm in} = r_{\rm isco}$ and the outer edge of disk is at $r_{\rm out} = 20M$.}
\label{image1}
\end{figure}

\begin{figure}[H]
\centering
\includegraphics[width=3.0in]{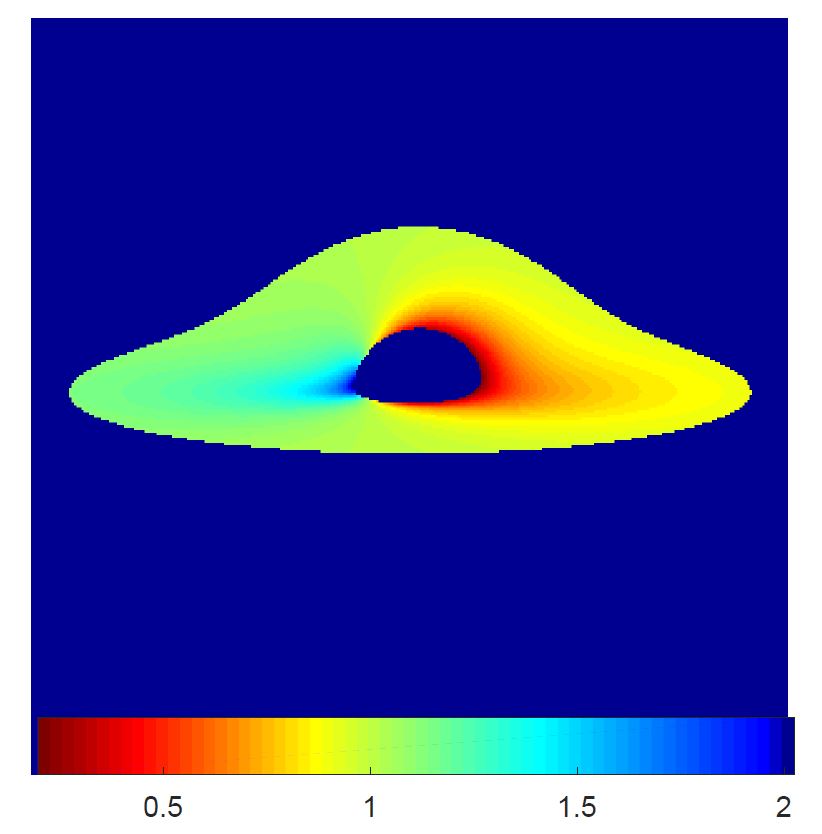}
\includegraphics[width=3.0in]{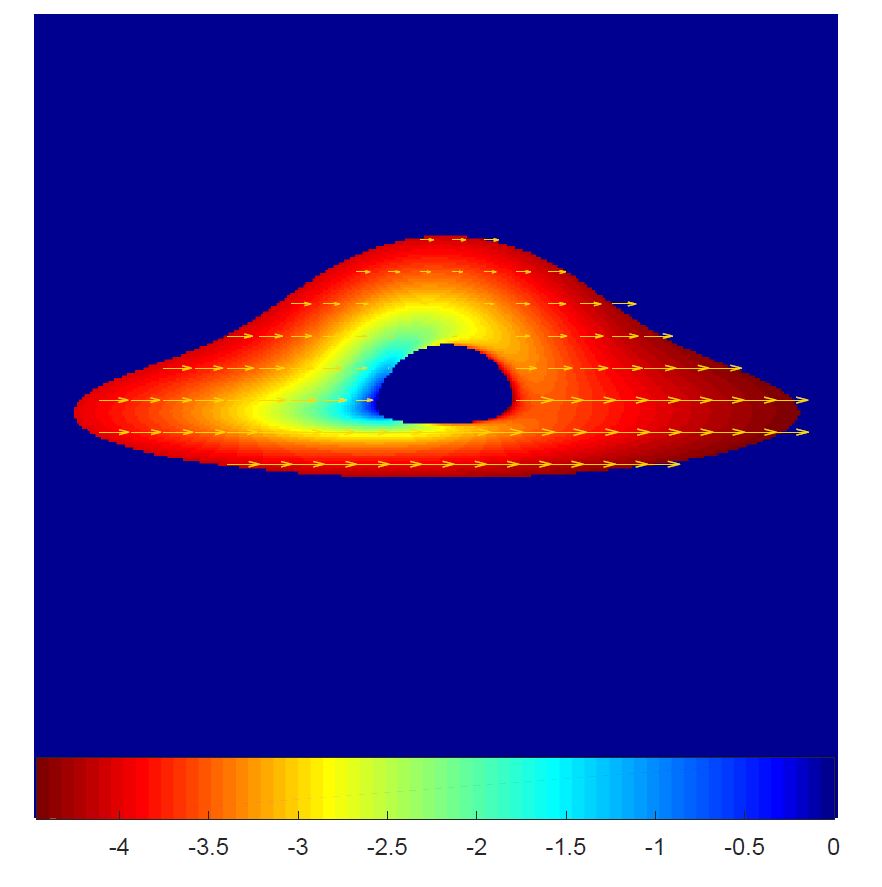}
\caption{\footnotesize Ray-traced redshifted image (left panel) and intensity and polarization profile (right panel) of a lensed accretion disk around rotating hairy Horndeski BH. The inclination angle is set to $i=80^{\circ}$, spin parameter is $a=0.6M$ and hair parameter is $h=-0.8M$ with $M = 1$. The inner edge of disk is at $r_{\rm in} = r_{\rm isco}$ and the outer edge of disk is at $r_{\rm out} = 20M$.}
\label{image2}
\end{figure}

\begin{figure}[H]
\centering
\includegraphics[width=3.0in]{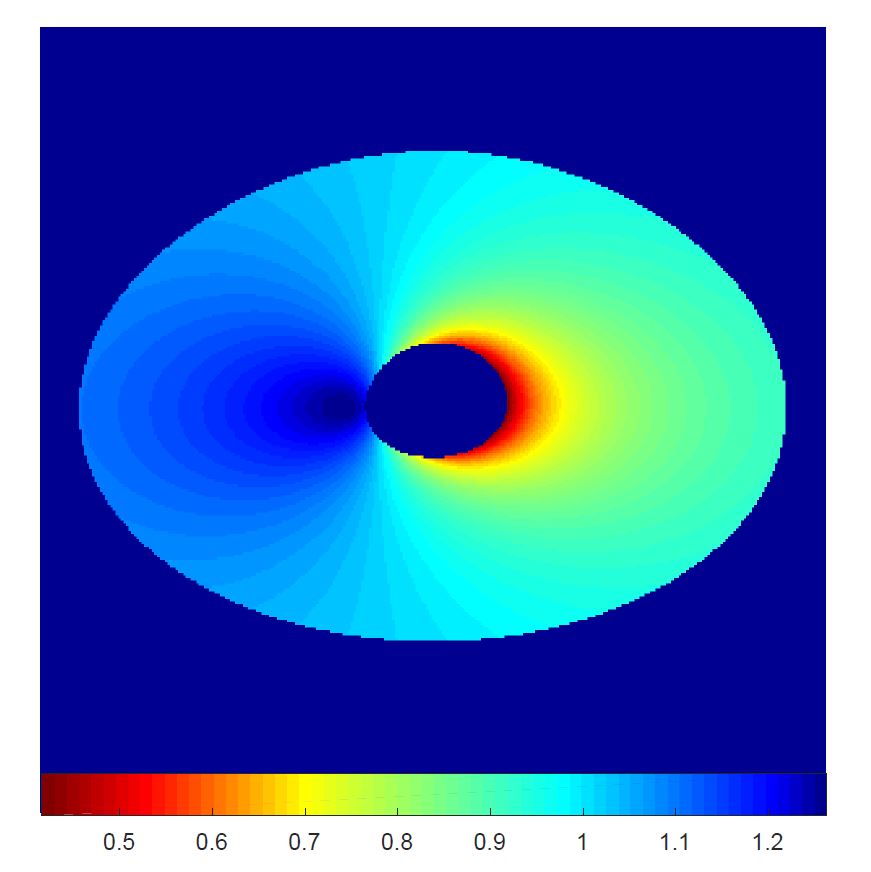}
\includegraphics[width=3.0in]{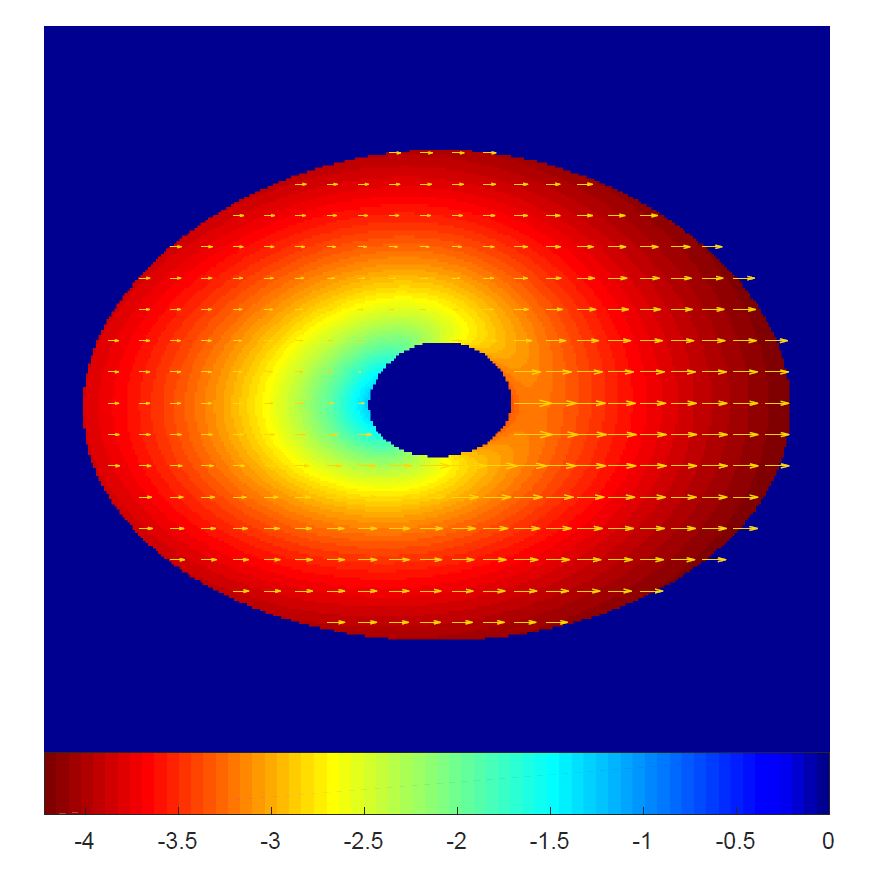}
\caption{\footnotesize Ray-traced redshifted image (left panel) and intensity and polarization profile (right panel) of a lensed accretion disk around rotating hairy Horndeski BH. The inclination angle is set to $i=50^{\circ}$, spin parameter is $a=0.6M$ and hair parameter is $h=-0.8M$ with $M = 1$. The inner edge of disk is at $r_{\rm in} = r_{\rm isco}$ and the outer edge of disk is at $r_{\rm out} = 20M$.}
\label{image3}
\end{figure}

Both relativistic effects including the gravitational redshift due to the existence of the BH and the relativistic Doppler effect due to the rotation of the disk lead to the changes of the observed image, as shown in the Fig.~\ref{image1}-Fig.~\ref{image3}. The figures show the Doppler blueshift in the left half-part of the plane which can exceed the overall gravitational redshift (left panels). Also, the intensity distribution is concentrated on a small region near the BH on the left side of accretion disk (right panels).

In Fig.~\ref{image1} and Fig.~\ref{image2}, we plot the ray-traced redshifted image (left panel) and also the intensity and polarization profile of an accretion disk (right panel) around a Kerr BH  with $h=0$ and a rotating hairy Horndeski BH for $h=-0.8M$, respectively. We set the spin parameter $a=0.6M$, $M=1$, and the inclination angle $i=80^{\circ}$. Fig.~\ref{image3} is similar to Fig.~\ref{image2} but presented at the arbitrary inclination angle $i=50^{\circ}$. The color bar shows the degree of the redshift of radiant light emitted from the accretion disk in ray-traced redshifted images, and the intensity of radiation on the accretion disk for intensity and polarization profiles. Also, we note that the size and the direction of polarization are shown by the length and direction of arrow in the right panel of figures.

The ray-traced redshift images are dependent on the hair parameter $h$ and the inclination angle of the accretion disk $i$. Comparing Fig.~\ref{image1} and Fig.~\ref{image2} shows that the region of the central shadow increases with the decrease of $|h|$. Also, the size and orientation of polarization vectors in the right panels of the figures decrease with increasing $|h|$, so that the Kerr space-time with $h=0$ has the most change of polarization vector in a small region on the accretion disk in the vicinity of BH. On the other hand, comparing Fig.~\ref{image2} and Fig.~\ref{image3} shows that the inclination angle has a notable effect on both the ray-traced redshifted image and the intensity and polarization profile of accretion disk, so that for the high inclination angles the blueshift in the left half of the plate increases.

\section{Conclusions}
\label{6-conclusions}
In the present work, we considered thin accretion disks around rotating hairy BHs in Horndeski gravity. First, we studied the geodesic motion of test particles in the space-time of these BHs and obtained the relevant quantities, namely the specific angular momentum $\tilde{L}$, specific energy $\tilde{E}$, angular velocity $\Omega$ and the ISCO radius for particles moving in circular orbits. We numerically obtained the ISCO radius for both spin, $a$, and the hair parameter, $h$. It is shown that by increasing both the spin parameter and the absolute value of the hair parameter, the ISCO radius decreases. Then, using the steady-state Novikov-Thorne model, we obtained the physical properties of accretion disks including the energy flux $F(r)$, temperature distribution $T(r)$, and the emission spectra $L(\nu)$ for some values of the hair parameter and plotted their profiles. The decrease of ISCO radius shows that the presence of scalar filed weakens the strength of gravitational field and thus the inner edge of disk shifts to lower and lower radii. Therefore, the disks around rotating BHs in Horndeski gravity are hotter and more luminous than the Kerr BH in the absence of hair.

Furthermore, we investigated the images of thin accretion disks around rotating hairy Horndeski BHs using a ray-tracing technique. We made the ray-traced redshifted images, intensity and polarization profiles of disks for both hairy Horndeski BH and the Kerr BH in the absence of hair. It is shown that with increasing the absolute value of the hair parameter $h$, the central shadow area decreases and the size and orientation of polarization vectors also decrease. Moreover, the results show that the inclination angle has a pronounced effect on the disk images.

\section*{Acknowledgments}
The work of Mohaddese Heydari-Fard is supported by the Iran National Science Foundation (INSF) and the Research Council of Shahid Beheshti University under research project No. 4016024.

\end{document}